# On a fusion chain reaction via suprathermal ions in high-density H-$^{11}$B plasma

## Fabio Belloni

*European Commission, Directorate-General for Research and Innovation, Euratom Research, Brussels, Belgium; email: fabio.belloni@ec.europa.eu*

**Abstract**

The $^{11}$B$(p,\alpha)2\alpha$ fusion reaction is particularly attractive for energy production purposes because of its aneutronic character and the absence of radioactive species among reactants and products. Its exploitation in the thermonuclear regime, however, appears to be prohibitive due to the low reactivity of the H-$^{11}$B fuel at temperatures up to 100 keV. A fusion chain sustained by elastic collisions between the $\alpha$ particles and fuel ions, this way scattered to suprathermal energies, has been proposed as a possible route to overcome this limitation. Based on a simple model, this work investigates the reproduction process in an infinite, non-degenerate H-$^{11}$B plasma, in a wide range of densities and temperatures which are of interest for laser-driven experiments ($10^{24} \lesssim n_e \lesssim 10^{28}$ cm$^{-3}$, $T_e \lesssim 100$ keV, $T_i \sim 1$ keV). In particular, cross section data for the $\alpha$-$p$ scattering which include the nuclear interaction have been used. The multiplication factor, $k_\infty$, increases markedly with electron temperature and less significantly with plasma density. However, even at the highest temperature and density considered, and despite a more than twofold increase by the inclusion of the nuclear scattering, $k_\infty$ turns out to be of the order of $10^{-2}$ only. In general, values of $k_\infty$ very close to 1 are needed in a confined scheme to enhance the suprathermal-to-thermonuclear energy yield by factors of up to $10^3 \div 10^4$.

**Keywords**

## 1. Introduction

The *advanced fusion fuel* (Dawson, 1981) based on a H-$^{11}$B mixture would exploit the reaction

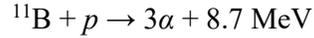

$$^{11}\text{B} + p \rightarrow 3\alpha + 8.7 \text{ MeV}$$

which is particularly attractive as it involves only abundant and stable isotopes in the reactants, and no neutron in the reaction products. The reaction cross section shows a main, wide resonance at 612 keV in the CM system (Fig. 1a), with a maximum value of 1.4 b (Sikora and Weller, 2016). A second, narrow resonance at lower energy (147 keV) peaks to about 0.1 b. Summed over the three reaction channels – the low-BR $^{12}$C$^*$ direct breakup and the sequential decays via $^8$Be$_{\text{gs}}$ or $^8$Be$^*$ (Becker *et al.*, 1987) – the energy spectrum of the generated $\alpha$ particles is a continuum which extends up to about 6.7 MeV in the lab, for a 675 keV incoming proton. The spectrum is strongly peaked around 4 MeV (Stave *et al.*, 2011).

A major drawback of H$^{11}$B fuel is due to the fact that the reaction cross section results in a much lower reactivity compared to conventional DT fuel, at the temperatures currently achievable in magnetically or inertially confined plasmas (of the order of 10 keV); Fig. 1b. This issue poses tremendous challenges to the feasibility of $p$-$^{11}$B fusion. Nevertheless, the fact that three charged, massive, energetic particles are produced in the reaction, suggests that the fusion yield could effectively be enhanced by a non-thermal effect induced by those particles, which is the elastic scattering of fuel ions to energies corresponding to the highest values of the fusion cross section (Brueckner and Brysk, 1973; Moreau, 1977). While thermalizing, some of the ions in these showers can undergo further fusion, eventually setting a chain reaction up. As a matter of fact, the fusion-born $\alpha$'s can easily be stopped in the plasma, especially at very high densities. At those densities, moreover, the $\alpha$'s tend to lose energy mostly to plasma ions rather than to electrons. This is certainly true in degenerate conditions (Gryzinski, 1958; Son and Fisch, 2004) and, away from degeneracy, e.g. for $T_e \gtrsim T_i/10$ when $T_i \gtrsim 150$ keV and the boron-to-hydrogen ion concentration is lower than 50% (Levush and Cuperman, 1982).

In an infinite, homogeneous H$^{11}$B plasma, an $\alpha$ particle emitted at a certain energy $E_{\alpha,0}$ is characterised by the multiplication (or reproduction) factor $k_\alpha(E_{\alpha,0})$, which is the average number of secondary $\alpha$-particles generated via suprathermal processes during its slowing down. For the purpose of this work, the multiplication factor in terms of fusion events, $k_\infty$, is well defined by the relation

$$k_\infty = \int k_\alpha(E_{\alpha,0})\varphi(E_{\alpha,0})dE_{\alpha,0} \qquad (1)$$

where $\varphi$ is the $\alpha$ spectral distribution normalised to 1. Each fusion event gives rise to new generations of events along a geometric progression. The number of events adds up, on average, to $1 + k_\infty + k_\infty^2 + \ldots + k_\infty^l + \ldots$, where $l$ is the number of generations. Along these lines, it is not difficult to show that upon a thermonuclear rate per unit volume $R$, the time evolution of the cumulative number density of fusion events, $n_f(t)$, is given by

$$n_f(t) = \frac{R\tau_g}{1 - k_\infty}\left[\frac{t}{\tau_g} + \frac{k_\infty}{\ln k_\infty}\left(1 - k_\infty^{t/\tau_g}\right)\right]$$

$$(2)$$

where $\tau_g$ is the average period between two consecutive generations, and the initial condition $n_f(0) = 0$ has been assumed. Depending on whether i) $k_\infty < 1$, ii) $k_\infty = 1$, or iii) $k_\infty > 1$, $n_f$ can respectively

i)      increase linearly with $t$, asymptotically to $Rt/(1 - k_\infty)$, for $t \to \infty$,

ii)     increase quadratically with $t$, as it reduces to $R\left(t + t^2/2\,\tau_g\right)$, or

iii)    diverge exponentially, with the growth rate $\ln k_\infty/\tau_g$.

It goes without saying that the capability to achieve a chain reaction with multiplicity $k_\infty \gtrsim 1$ would play a significant – if not indispensable – role in the possible exploitation of $H^{11}B$ fusion for energy production purposes, especially under laser-driven schemes for plasma generation and confinement (Hora *et al.*, 2015, 2016, 2017).

Early $k_\alpha$ calculations by Moreau (1976, 1977), however, returned values of the order of $10^{-2}$ in a plasma with $100 \leq T_e \leq 300$ keV, $T_i = 0$, and Coulomb logarithm $ln\,\Lambda = 5$. The boron-to-hydrogen ion concentration used is unclear (perhaps 8.9%, the optimum for thermonuclear ignition); moreover, only the Coulomb interaction was taken into account in the $\alpha$-ion scattering, neglecting the nuclear (hadronic) one, which can instead turn out to be predominant, as we shall see. Recently, Hora *et al.* (2015, 2016) and Eliezer *et al.* (2016) have claimed evidence of the chain reaction in experiments at the Prague Asterix Laser System (PALS), Czech Republic, where an unprecedented fusion yield ($4 \times 10^8$ $\alpha$'s per laser pulse) had been achieved by irradiation of an H-enriched, B-doped Si target (Picciotto *et al.*, 2014; Margarone *et al.*, 2015). On the contrary, Shmatov (2016) and Belloni *et al.* (2018) have refuted this possibility on the basis of stopping power and $\alpha$-$p$ collision rate arguments. Lately, Eliezer and Martinez-Val (2020) have proposed the concept of a possible $H^{11}B$ fusion reactor where the stopping power problem is circumvented by the application of an external electric field.

As it is evident, a systematic study of the chain reaction multiplicity in high-density $H^{11}B$ plasma has been missing so far. Studies of this kind exist for DT fuel (e.g. Brueckner and Brysk, 1973; Peres and Shvarts, 1975; Afek *et al.*, 1978; Kumar *et al.*, 1986) and will briefly be outlined in Sect. 4. Prompted by recent theoretical and experimental advances in laser-driven $p$-$^{11}B$ fusion (Hora *et al.*, 2017; Giuffrida *et al.*, 2020), the advent of ultra-high-power (tens PW) laser systems as well as the growing interest of the scientific community in the potentiality of $H^{11}B$ fuel for energy production (Rostoker *et al.*, 1997; Belyaev *et al.*, 2005; Labaune *et al.*, 2013; Cowen, 2013), this work reports an analysis of the multiplication process in a wide range of plasma densities and temperatures which are of interest for current and future laser-based experiments. In particular, cross section data for the $\alpha$-$p$ scattering which include the nuclear interaction have been used. The aspiration is that results and conclusions can help inform the choice of parameters and the development of techniques in future experiments, with a view to maximising multiplication effects.

## 2. Theory

We assume a two-temperature $(T_e, T_i)$, H$^{11}$B plasma with $n_j$ ions per unit volume (the subscript $j$ stands for $p$ or B); the density ratio $n_B/n_p$ is denoted by $\gamma$. Indicating by $E_{j,0}$ the energy of an ion just after the scattering by an $\alpha$ particle, we will consider as suprathermal those secondary ions for which $E_{j,0} \gg T_i$. In this limit, the number of ions of the species $j$ scattered into the energy interval $(E_{j,0}, E_{j,0} + dE_{j,0})$ through the path length $dx$ of the $\alpha$ particle is given by

$$d^2 N_j = n_j \sigma_s (E_\alpha, E_{j,0}) dE_{j,0} dx \tag{3}$$

where $\sigma_s$ is the differential scattering cross section. The spectral distribution of these ions through the entire path length of the $\alpha$ particle is then

$$\frac{dN_j}{dE_{j,0}} (E_{j,0}; E_{\alpha,0}) = n_j \int_{\frac{3}{2}T_i}^{E_{\alpha,0}} \sigma_s (E_\alpha, E_{j,0}) \left( \frac{dE_\alpha}{dx} \right)^{-1} dE_\alpha \tag{4}$$

where $dE_\alpha/dx$ is the stopping power of the $\alpha$ particle (taken as a positive quantity). For pure Coulomb scattering, $\sigma_s$ is given by the well-known Rutherford cross section, $\sigma_R$, which reads, in terms of $E_{j,0}$, as

$$\sigma_R(E_\alpha, E_{j,0}) = \frac{\pi(z_\alpha z_j e^2)^2}{E_\alpha E_{j,0}^2} \frac{m_\alpha}{m_j} H(E_{j,0}^{max} - E_{j,0})$$

(5)

where the $z$'s are particle electric charges in units of the elementary charge $e$, the $m$'s are particle masses, $H$ is the Heaviside step function, and the endpoint energy $E_{j,0}^{max}$ is given, from basic kinematics, by

$$E_{j,0}^{max} = \eta_{\alpha j} E_\alpha$$

(6)

with $\eta_{\alpha j} = 4m_\alpha m_j/(m_\alpha + m_j)^2$.

Denoting by $k_{\alpha j}$ the contribution to $k_\alpha$ of the suprathermal population of the species $j$, it is straightforward to see that for an $\alpha$ particle of energy $E_{\alpha,0}$, the spectrum of $k_{\alpha j}$ over $E_{j,0}$ is linked to the spectrum of $N_j$ by

$$\frac{dk_{\alpha j}}{dE_{j,0}}(E_{j,0}; E_{\alpha,0}) = 3 \, P_j(E_{j,0}) \frac{dN_j}{dE_{j,0}}(E_{j,0}; E_{\alpha,0})$$

(7)

where $P_j$ is the fusion probability of the ion throughout its thermalisation, and the factor of 3 is the number of $\alpha$'s per fusion event. $k_{\alpha j}(E_{\alpha,0})$ is calculated by numerical integration of Eq. (7) over $E_{j,0}$. In turn, $k_\alpha$ is calculated by summing over the contribution of each ion species, i.e.

$$k_\alpha(E_{\alpha,0}) = k_{\alpha p}(E_{\alpha,0}) + k_{\alpha B}(E_{\alpha,0})$$

(8)

Finally, $k_\infty$ is calculated from $k_\alpha(E_{\alpha,0})$ through Eq. (1). By the same way, it is also meaningful to calculate the spectrum $dk_{\infty,j}/dE_{j,0}$ from $dk_{\alpha j}/dE_{j,0}$.

Concerning the fusion probability in Eq. (7), it is easy to show (e.g. Giuffrida *et al.*, 2020) that in the cold-ion approximation for the target species, and for $P_j \ll 1$ (which is our case, see Sect. 3), the following relation holds

$$P_{p(B)}\big(E_{p(B),0}\big) = n_{B(p)} \int_0^{E_{p(B),0}} \sigma_f(E_{CM}) \big(\frac{dE_{p(B)}}{dx}\big)^{-1} dE_{p(B)}$$

(9)

where $dE_j/dx$ is the stopping power, $\sigma_f$ is the fusion cross section, and $E_{CM}$ is the CM energy of the $p$-$^{11}$B system, i.e.

$$E_{CM} = \frac{m_{B(p)}}{m_p + m_B} E_{p(B)}$$

(10)

Calculations of the abovementioned quantities have been performed in conditions relevant to laser-driven fusion plasmas and are reviewed in Sect. 3. In computations, we have adopted the following input data or models and considerations.

*α spectrum, $\varphi\big(E_{\alpha,0}\big)$*

By the sake of simplicity, we have used the crude two-group approximation according to which, on average, one $\alpha$ particle is emitted at energy $E_{\alpha,I} = 1$ MeV and the other two at $E_{\alpha,II} = 4$ MeV (Stave *et al.*, 2011). Consequently, $\varphi\big(E_{\alpha,0}\big)$ in Eq. (1) is given by

$$\varphi\big(E_{\alpha,0}\big) = \frac{1}{3}\big[\delta\big(E_{\alpha,0} - E_{\alpha,I}\big) + 2\delta\big(E_{\alpha,0} - E_{\alpha,II}\big)\big]$$

(11)

where $\delta$ is the Dirac delta function.

*Fusion cross section, $\sigma_f$*

The analytic approximation of Nevins and Swain (2000) has been used below 3.5 MeV, and an interpolation of TENDL evaluated data (Koning *et al.*, 2019) at higher energies (Fig. 1a).

*Elastic scattering cross section, $\sigma_s$*

For the $\alpha$-$p$ scattering, evaluated cross section data have been interpolated from SigmaCalc (Gurbich, 2016). A comparison with $\sigma_R$ is displayed in Fig. 2, which shows that the nuclear contribution dramatically increases the cross section, within a factor of 3 for $E_\alpha \lesssim 2$ MeV, up to a factor of 10 around 4 MeV and at high values of $E_{p,0}$, and by hundreds of times at progressively higher values of $E_\alpha$ and $E_{p,0}$. Nevertheless, for $E_\alpha > 4$ MeV and $E_{p,0} < 1$ MeV, a wide area exists where $\sigma_s/\sigma_R < 1$, which is an effect of the interference between the Coulomb and nuclear scattering amplitudes (Perkins and Cullen, 1981). For the $\alpha$-$^{11}$B scattering, the Rutherford cross section has been used as the higher Coulomb barrier makes the nuclear contribution negligible at the energies concerned.

*Stopping powers, $dE_\alpha/dx$ and $dE_j/dx$*

The Spitzer-Sivukhin model (Spitzer, 1956; Sivukhin, 1966) has been used, in the form for a multicomponent (electrons and ion species), two-temperature H$^{11}$B plasma detailed in Levush and Cuperman (1982). Earlier, this model had also been used by Moreau (1976, 1977) for slowdown calculations in high-density H$^{11}$B plasma. With a view to the subsequent discussion, it is worth recalling the form of the electronic stopping power in the expressions of $dE_\alpha/dx$ and $dE_j/dx$, i.e.

$$dE_{qe}/dx = n_e ln\Lambda_{qe}(n_e,T_e)g_{qe}(E_q,T_e) \qquad (12)$$

where the subscript $q$ stands for $\alpha$, $p$ or B, and the functions $ln\Lambda_{qe}$ and $g_{qe}$ are given by Sivukhin (1966). Analogous formulas hold for the $q$-$p$ and $q$-$^{11}$B components of the stopping power.

*Plasma electron density, $n_e$*

With reference to Eqs. (4) and (9), it is useful to write the ion densities appearing therein in terms of the electron density, $n_e$, and $\gamma$, i.e.

$$n_p = n_e/(z_p + z_B\gamma) \tag{13a}$$

$$n_B = n_e\gamma/(z_p + z_B\gamma) \tag{13b}$$

Orders of magnitudes between $10^{24}$ and $10^{28}$ cm$^{-3}$ have been considered for $n_e$. As a term of reference, for amorphous boron in STP conditions, $n_B = 1.3 \times 10^{23}$ cm$^{-3}$, hence $n_e = 6.5 \times 10^{23}$ cm$^{-3}$.

*Boron-to-hydrogen ion concentration, $\gamma$*

If one considers, by the sake of simplicity, only the electronic stopping power in the expressions of $dE_\alpha/dx$ and $dE_j/dx$ [Eq. (12)], it is straightforward to note that the apparent linear dependence on $n_e$ in Eqs. (4) and (9) actually cancels out, leaving factors which depend on $\gamma$ according to Eqs. (13). This implies that Eq. (7) and derived quantities depend on $\gamma$ through the overall factor $\gamma/(z_p + z_B\gamma)^2$, which has a maximum at $\gamma = 0.2$ for $z_p = 1$ and $z_B = 5$. In reality, the dependence of the multiplication factors on $\gamma$ is obviously much more complicate because of the dependence on $n_j$ of the ion-ion components of the stopping powers. The optimum $\gamma$ has to be calculated numerically and depends, moreover, on the specific set of parameters entering the equations. Its value and the corresponding maximum values of $k_\alpha$ and $k_\infty$, however, are in general not dramatically affected, as shown in Sect. 3.

*Temperatures, $T_e$ and $T_i$*

Minimising $dE_\alpha/dx$ and $dE_j/dx$ in Eqs. (4) and (9), respectively, requires high values of both $T_e$ and $T_i$ in a classic (Maxwell-Boltzmann) plasma. As a precondition, one wants to deal with a fully ionized plasma in order to reduce the electronic component of the stopping power (Giuffrida *et al.*, 2020); accordingly, $T_e$ should be much higher than the ionization energy of $B^{4+}$, which is 0.34 keV for the isolated ion (Lide, 2000) and less in high-density matter (More, 1993). Even when $T_e$ is higher than the $B^{4+}$ ionization energy, electrons can still be Fermi-degenerate at the high densities considered here. The Fermi energy $E_F$, which scales as $n_e^{2/3}$, has been plotted in Fig. 3 for reference. In view of further reducing the electronic stopping power, we have verified that in our density domain it is convenient to work at low degeneracy, e.g. at $T_e/E_F > 5$, compared to the fully degenerate case (Son and Fisch, 2004; Giuffrida *et al.*, 2020). Accordingly, at a given $n_e$, we have chosen to work with a classic plasma with $T_e > 5E_F$, a condition which in our case ensures both full ionization and low degeneracy. On the other hand, at a given $T_e$, the lower $T_i$ the more effective is the suprathermal energy transfer. Indeed, in the limit $T_i = 0$, all the scattered ions are obviously suprathermal; moreover, from basic kinematics, the collisional energy transfer from the $\alpha$ particle to the ion occurs as long as the velocity of the former is higher than that of the latter. We have chosen to set $T_i = 1$ keV, which is a good compromise among the needs to reduce the ion-ion component of the stopping power, increase the $\alpha$-to-ion energy partition, and ensure a suprathermal spectrum as wide as possible ($T_i \ll E_{j,0} \le E_{j,0}^{max}$). As a matter of fact, the contribution to $k_{\alpha p}$ of protons with $E_{p,0} < 10$ keV is absolutely negligible; see Sect. 3. Incidentally, in a low-$T_i$ scheme the thermonuclear burn will be very modest and will just be used to seed the chain reaction, which is expected to provide most of the energy output. The choice of parameters $10^{24} \le n_e \le 10^{28}$ cm$^{-3}$, $T_i = 1$ keV, $max[T_i, 5E_F(n_e)] \le T_e \le 100$ keV (Fig. 3) appears to be of interest for present and future laser-driven experiments where hot electrons are generated in shocked or inertially compressed targets; see e.g. the recent experiment of Giuffrida *et al.* (2020) on a shocked hydrogen-rich boron nitride target, where $n_e \sim 10^{24}$ cm$^{-3}$ and $T_e \sim T_i \sim 1$ keV have been estimated.

## 3. Results

The $^{11}$B-ion contribution to $k_\alpha$ in Eq. (8) turns out to be in any case much smaller than that of the proton, by a factor of at least 100, as we have verified by direct computation for values of $E_{\alpha,0}$ up to 10 MeV. Accordingly, explicit results for suprathermal $^{11}$B ions will not be reported in the followings. The physical reasons for the negligible role of $^{11}$B recoils in the multiplication process will be outlined in Sect. 4. In the present Section, focus is made on the results of calculations for $P_p(E_{p,0})$, $dN_p/dE_{p,0}$, $dk_{\alpha p}/dE_{p,0}$, $dk_{\infty,p}/dE_{p,0}$, $k_{\alpha p}(E_{\alpha,0})$, and $k_\infty$.

The fusion probability of Eq. (9) has been plotted in Fig. 4 as a function of proton energy for representative values of $n_e$ and $T_e$, with $T_i = 1$ keV and $\gamma = 0.2$. The shape of the curves resembles the main features of $\sigma_f$. The curves, in particular, exhibit marked knees at energies around the main cross section resonances. Further, the fusion probability becomes negligibly small below 100 keV. Within our domain of parameters, values of $P_p$ of the order of $10^{-3}$ are attained at proton energies between approximately 400 and 500 keV. The fusion probability is enhanced more effectively by increasing $T_e$ rather than $n_e$ (the curves at $10^{25}$ and $10^{26}$ cm$^{-3}$ are almost indistinguishable); furthermore, the $T_e$-driven enhancement is amplified by proton energy. While the dependence on $T_e$ is entirely due to the stopping power in Eq. (9), the slight dependence on $n_e$ results from its quasi cancellation in the product between $n_B$ and $\left(dE_p/dx\right)^{-1}$, as discussed in Sect. 2.

The spectral analysis of scattered protons and multiplication factors is provided in Fig. 5. In panel a), $dN_p/dE_{p,0}$ has been plotted according to Eq. (4), for the two reference values of $E_{\alpha,0}$ (viz. $E_{\alpha,I}$ and $E_{\alpha,II}$) and the elastic cross sections $\sigma_s$ and $\sigma_R$, respectively. First, one notes that the spectrum endpoint energies, as given by Eq. (6), are quite high, approaching 600 keV in the case of $E_{\alpha,I}$ and overcoming 2 MeV in the case of $E_{\alpha,II}$. Secondly, at the given values of $E_{\alpha,0}$, the trend of the curves corresponding to $\sigma_s$ and $\sigma_R$ is similar. Nevertheless, at values of $E_{p,0}$ of the order of 10 keV and beyond, one notes that the yield of scattered protons is significantly higher when the nuclear interaction is taken into account. At intermediate energies, of the order of 100 keV, the difference reduces significantly. At energies of the order of 1 MeV, the scattered yield is again enhanced by the nuclear interaction. Such "oscillation" in this proton energy range is basically due to the interference between the scattering amplitudes of the nuclear and Coulomb potentials, as mentioned in Sect. 2. In panel b), $dk_{\alpha p}/dE_{p,0}$ has been plotted according to Eq. (7), for the two reference values of $E_{\alpha,0}$ and the same set of parameters indicated in panel a). The distribution $dk_{\infty,p}/dE_{p,0}$ resulting from the $\alpha$ spectrum of Eq. (11) is also shown. Despite the increasing yield of scattered protons at low $E_{p,0}$, it is immediate to recognise that the contribution to $k_{\alpha p}$ and $k_{\infty,p}$ from protons with $E_{p,0} \lesssim$ 100 keV is negligible. This is obviously due to the extremely low fusion probability at those energies. One can also notice that the contribution to $k_{\infty,p}$ of the low-energy $\alpha$'s (blue curve) is rather limited.

A detailed analysis of the behaviour of $k_{\alpha p}(E_{\alpha,0})$ with $T_e$ and $n_e$ is shown in Fig. 6. In panel a), curves have been generated for three different values of $T_e$ (10, 50 and 100 keV), keeping $n_e$ fixed at $10^{26}$ cm$^{-3}$. As a term of comparison, a curve based only on the Rutherford $\alpha$-$p$ scattering has been calculated at $T_e = 50$ keV. In panel b), $n_e$ has been varied by 1-decade steps from $10^{24}$ to $10^{28}$ cm$^{-3}$ while keeping $T_e$ fixed at 100 keV. In all cases, the curves quickly drop below $E_{\alpha,0} \simeq 2$ MeV. Above 4 MeV, their shape is approximately linear in the semi-log plot, meaning an exponential increase with $E_{\alpha,0}$. A fit with a function of the form $k_{\alpha p} \propto exp(aE_{\alpha,0})$ on the curves of panel b) returns $a \simeq 0.45$ MeV$^{-1}$ for $4 \leq E_{\alpha,0} \leq 10$ MeV, meaning that $k_{\alpha p}$ increases by a factor of about 2.5 each time $E_{\alpha,0}$ increases by 2 MeV.

At a given $E_{\alpha,0}$, $k_{\alpha p}$ increases with both $T_e$ and $n_e$, as expected from stopping power considerations. The slope of the straight portion of the curves in Fig. 6 slightly increases with $T_e$, whereas it is unaffected by variations of $n_e$. In the latter case, the spacing between adjacent curves increases with regularity upon tenfold increments of $n_e$. However, the logarithmic sensitivity of $k_{\alpha p}$ to $n_e$ is very limited; for instance, $\partial ln k_{\alpha p}/\partial ln n_e < 0.12$ at $T_e = 100$ keV, whereas for $n_e = 10^{26}$ cm$^{-3}$, $\partial ln k_{\alpha p}/\partial ln T_e$ approximately ranges from 2.3 at $T_e = 10$ keV to 1.8 at $T_e = 100$ keV. We also remark the effect of the use of $\sigma_s$ instead of $\sigma_R$ in the $\alpha$-$p$ scattering. In Fig. 6a, the gap of the respective curves at $T_e = 50$ keV progressively increases with $E_{\alpha,0}$, resulting in a value of $k_{\alpha p}$ which e.g., at $E_{\alpha,0} = 10$ MeV, is about 8 times higher when the nuclear interaction is taken into account. Despite this and the abovementioned favourable features, even at the highest values of $E_{\alpha,0}$, $T_e$ and $n_e$ considered in this work, $k_{\alpha p}$ remains significantly lower than 1.

Finally, the dependence on $\gamma$ of $k_{\alpha p}(E_{\alpha,I})$, $k_{\alpha p}(E_{\alpha,II})$ and the resulting $k_\infty$ is shown in Fig. 7, in the range $0 \leq \gamma \leq 1$ and for the most multiplication-effective (viz the highest) values of $n_e$ and $T_e$ explored in this work. As it is obvious, the curves vanish for $\gamma \to 0$ (too few $^{11}$B ions for the fusion reaction to occur) and decrease smoothly to 0 for $\gamma > 1$ (too few protons available for scattering). In between, a maximum occurs at values of $\gamma$ which depend on $E_\alpha$ and are however not far from 0.2, the value argued in Sect. 2 and used in the calculations above. More importantly, the differences in $k_{\alpha p}(E_{\alpha,I})$, $k_{\alpha p}(E_{\alpha,II})$ and $k_\infty$ between the case $\gamma = 0.2$ and the respective optimal $\gamma$'s are limited to the order of 10%. We remark that in the conditions of Fig. 7, the peak value for our estimate of $k_\infty$ is only of the order of $10^{-2}$.

## 4. Discussion

As anticipated, it is instructive remarking the physical reasons behind the very modest impact of suprathermal $^{11}$B ions on the multiplication process. It is already evident from Eq. (10) for the $p$-$^{11}$B CM energy in the fixed-target reaction that the fusion probability of suprathermal $^{11}$B ions is much smaller than that of protons, at the same particle energy. Indeed, $E_{CM}$ is suppressed by a factor $m_p/m_B$, resulting in very low values of $\sigma_f$ in Eq. (9). Moreover, at the same particle energy, $dE_B/dx$ is larger than $dE_p/dx$, which still depresses the integrand in Eq. (9). On the opposite, at given $E_{\alpha,0}$ and ion energy, $dN_B/dE_{B,0}$ tends to be larger than $dN_p/dE_{p,0}$, by a factor $\left(z_B/z_p\right)^2 m_p/m_B = 2.3$ when $\sigma_s = \sigma_R$ is assumed in Eq. (4). Also, the $^{11}$B suprathermal spectrum is slightly wider than the proton one, since $E_{B,0}^{max}/E_{p,0}^{max} = \eta_{\alpha B}/\eta_{\alpha p} = 1.2$. Nevertheless, the net result from Eq. (7) is that the $^{11}$B contribution to $k_\alpha$ in Eq. (8) is in any case much smaller than the proton one.

Though the findings of Sect. 3 prevent the possibility of achieving the chain reaction in realistic conditions, it is of upmost importance to study how and how much a weak multiplication regime, i.e. when $k_\infty < 1$ (and especially $k_\infty \ll 1$), can enhance the pure thermonuclear burn. In this respect, the ratio of the energy per unit volume produced during the confinement time $\tau_c$, $\mathcal{E}$, to the energy stemming from the sole thermonuclear burn, $\mathcal{E}_{th}$, is just $n_f(\tau_c)/R\tau_c$, where $n_f(t)$ is given by Eq. (2). We prefer to study this ratio in the form of the fractional increment $I \equiv (\mathcal{E} - \mathcal{E}_{th})/\mathcal{E}_{th}$, which is equivalent to the suprathermal-to-thermonuclear energy yield, $\mathcal{E}_{st}/\mathcal{E}_{th}$, since the suprathermal energy component, $\mathcal{E}_{st}$, is obviously $\mathcal{E} - \mathcal{E}_{th}$. Explicitly,

$$I(k_\infty) = \frac{\mathcal{E}_{st}}{\mathcal{E}_{th}}(k_\infty) = \frac{k_\infty}{1 - k_\infty}\left(1 + \frac{\tau_\alpha}{\tau_c}\frac{1 - k_\infty^{\tau_c/\tau_\alpha}}{ln k_\infty}\right)$$

(14)

where $\tau_\alpha$ is the spectrum-averaged thermalisation time of the $\alpha$'s, and we have used the fact that $\tau_g \approx \tau_\alpha$. Indeed, if one estimates $\tau_g$ as the average extinction time of the $\alpha$-induced recoil shower, then $\tau_g^2 \approx \tau_\alpha^2 + \tau_p^2$, where $\tau_p$ is the spectrum-averaged thermalisation time of the secondary protons; $\tau_p$ is obviously longer than the average thermalisation time of B ions, but much shorter than $\tau_\alpha$.

One immediately notes that $I$ depends on the ratio $\tau_c/\tau_\alpha$ as a parameter. At the plasma densities considered here and for $T_e \sim 5E_F$, $\tau_\alpha$ generally turns out to be of the order of 1 ps or lower. With $\tau_c \sim 1$ ns, $\tau_c/\tau_\alpha$ can then reach the order of $10^3$ or $10^4$. Notice that for a self-sustaining chain reaction (i.e. $k_\infty \geq 1$), $\tau_c/\tau_\alpha$ represents the maximum possible number of $\alpha$-particle generations within the time $\tau_c$.

Plots of $I$ as a function of $k_\infty$ are shown in Fig. 8, for several orders of magnitude of $\tau_c/\tau_\alpha$ and $k_\infty < 1$. In the limit $\tau_c/\tau_\alpha \to \infty$, Eq. (14) yields the asymptotic behaviour $I \sim k_\infty/(1 - k_\infty)$ which, being independent of $\tau_c/\tau_\alpha$, explains the saturation of the curves observed at high values of the parameter. For $k_\infty \ll 1$, one recognises the approximate scaling $I \sim k_\infty$ in Fig. 8. This means that in the plasma conditions investigated in this work, the burn enhancement due to the multiplication is of the order of 1% at most.

In the limit $k_\infty \to 1$, Eq. (14) yields $I \to (1/2)\,\tau_c/\tau_\alpha$. This opens the possibility of very large increments in the energy output (and consequently, high fusion gains); however, at high $\tau_c/\tau_\alpha$, $I$ raises steeply when $k_\infty \to 1$, so that $k_\infty$ has to lie very close to 1 to allow increments of the order of $\tau_c/\tau_\alpha$ being approached (e.g. $I \approx 0.37\,\tau_c/\tau_\alpha$ when $1 - k_\infty = \tau_\alpha/\tau_c$, for $\tau_c/\tau_\alpha \gg 1$).

To summarise, our parametric analysis has shown that $k_\infty$ increases markedly with $T_e$ and less significantly with $n_e$, with the optimum $\gamma$ lying between 0.2 and 0.4. The achievable fusion energy is further enhanced by the ratio $\tau_c/\tau_\alpha$. In the weak chain, however, the enhancement is quite limited moving from $\tau_c/\tau_\alpha \sim 10$ to $\tau_c/\tau_\alpha \sim 100$ while $k_\infty \lesssim 0.5$, and negligible beyond $\tau_c/\tau_\alpha \sim 100$ while $k_\infty \lesssim 0.9$. We note, moreover, that there is a trade-off between the requirements for rising $T_e$ up on the one hand, and keeping $\tau_c/\tau_\alpha$ sufficiently large on the other hand, since $\tau_\alpha$ also increases with $T_e$ (on the contrary, $\tau_\alpha$ decreases with $n_e$ as $1/n_e ln\Lambda$). For typical confinement times, however, values of $\tau_c/\tau_\alpha$ larger than at least 10 appear to be always ensured.

We conclude this Section by making contact with previous representative findings for DT fuel. Peres and Shvarts (1975) have shown that a chain reaction via elastic recoils can proceed in a cold infinite DT plasma at densities above $8.4 \times 10^{27}$ ions/cm$^3$. The optimum isotopic ratio is $n_D/n_T = 0.72$. In the analysis, they have also considered recoil-induced DD and TT fusion reactions and the scattering by their products. The main contributor to the chain reaction turns out to be the DT-born 14.1 MeV neutron, while the 3.5 MeV $\alpha$ particle contributes only a few percent. Indeed, if the neutron is disregarded, the medium is not critical even at $10^{29}$ ions/cm$^3$, the highest density considered by the authors. This observation is of interest for inertial confinement experiments, where the neutron can easily escape the compressed pellet. For a finite-temperature, infinite plasma, however, Afek *et al.* (1978) have found lower critical densities; for instance, $1.0 \times 10^{27}$ ions/cm$^3$ at $T_e = T_i = 14$ keV and $n_D/n_T = 0.64$. For the finite-temperature, finite-size case, Kumar *et al.* (1986) have estimated an upper bound of 0.5 for the suprathermal fusion probability associated to the DT neutron in a pellet compressed to a thickness ($\rho r$) of a few g/cm$^2$, at a density of $6.0 \times 10^{25}$ ions/cm$^3$ (roughly 1000 times the solid density), $T_e = T_i = 40$ keV, and $n_D / n_T = 1$. We conclude that the suprathermal contribution to the fusion yield is substantially lower in H$^{11}$B fuel compared to DT fuel, in similar plasma conditions and for cases of practical interest.

### 5. Conclusion

We have investigated the possibility of a fusion chain reaction via $\alpha$-recoiled ions in high-density, non-degenerate H$^{11}$B plasma, under conditions which are of interest for laser-driven experiments ($10^{24} \lesssim n_e \lesssim 10^{28}$ cm$^{-3}$, $T_e \lesssim 100$ keV, $T_i \sim 1$ keV). On the basis of a simple model, the multiplication factor for individual fusion events ($k_\infty$) has been estimated in terms of the energy-dependent multiplication factor for individual $\alpha$ particles ($k_\alpha$), by averaging over the $\alpha$ emission spectrum. A spectral analysis of the suprathermal proton population and of the multiplication factors is also reported.

We have found that the contribution of suprathermal $^{11}$B ions to $k_\alpha$ and $k_\infty$ is of the order of 1% only. In the case of the scattered proton, the complete elastic cross section, accounting also for the nuclear interaction, must be used in calculations. For instance, the value of $k_\alpha$ for the most probable $\alpha$ emission energy (about 4 MeV) turns out to be more than twice that found for a pure Coulomb scattering. The spectral analysis shows that only protons with recoil energies higher than or comparable to 100 keV contribute to the multiplication factors. This important limitation is essentially due to the drop of the fusion probability at lower energies.

The parametric analysis shows that $k_\alpha$ increases with both $T_e$ and $n_e$, though it is much more sensitive to $T_e$. The optimum $\gamma$ lies between 0.2 and 0.4. In general, $k_\alpha$ quickly drops below $E_{\alpha,0} \simeq 2$ MeV, while it increases nearly exponentially above 4 MeV, up to at least 10 MeV, the highest $\alpha$ energy considered in this work. Even for the highest values of $n_e$ and $T_e$ considered, $k_\alpha$ (hence, $k_\infty$) remain significantly lower than 1; $k_\alpha = 0.2$ for $E_{\alpha,0} = 10$ MeV, and $k_\infty \approx 0.01$. While $k_\infty \lesssim 0.3$, the fractional increment in the energy output relative to the thermonuclear burn scales linearly with $k_\infty$, being practically insensitive to the parameter $\tau_c/\tau_\alpha$ and remaining, therefore, quite limited. On the contrary, for $k_\infty \to 1$, the burn enhancement approaches the order of magnitude of $\tau_c/\tau_\alpha$, which can easily be made as large as $10^3$ or $10^4$ in experiments.

Increasing $k_\infty$ above the order of $10^{-2}$, however, appears problematic in realistic laser-driven plasma conditions, meaning those presently achievable or likely to be achieved in the near future. One notes, moreover, that $k_\infty$ in $H^{11}B$ fuel is substantially lower than in DT fuel, *ceteris paribus*. Novel ideas are needed in order to exploit the full potential of suprathermal multiplication in $H^{11}B$ fuel. For instance, several authors (Belyaev *et al.*, 2005; Picciotto *et al.*, 2014; Giuffrida *et al.*, 2020) have reported $\alpha$ spectra shifted towards higher energies (up to 10 MeV) in laser-driven $p$-$^{11}B$ fusion experiments. Giuffrida *et al*. have ascribed this phenomenon to the acceleration of the fusion-born $\alpha$'s by the same laser-induced electric field which accelerates the protons to MeV energies. With a proper choice of target characteristics and laser parameters, such an effect could represent an excellent means to exploit the nearly exponential increase of $k_\alpha$ at high $\alpha$ energy.

**Acknowledgements**

The author wishes to thank many of the participants in the 1[st] Discussion Workshop on Proton-Boron Fusion, 5 June 2020 (held remotely because of the Covid-19 outbreak), for debating the subject of this work during and after the event. The valuable support of D. Ostojić is also gratefully acknowledged. All views expressed herein are entirely of the author and do not, in any way, engage his institution.

**Figure captions**

**Fig. 1.** a) $p$-$^{11}$B fusion cross section as a function of the CM energy, based on the analytic approximation of Nevins and Swain (2000) below 3.5 MeV and, above, on TENDL evaluated data. b) Reactivity as a function of ion temperature for H$^{11}$B fuel and, as a term of comparison, DT fuel. Plots are based on the analytic approximations of Nevins and Swain (2000) and Bosch and Hale (1992), respectively.

**Fig. 2.** Logarithmic contour plot of the ratio $\sigma_S/\sigma_R$ in the $\left(E_\alpha, E_{p,0}\right)$ plane, with the boundary given by Eq. (6).

**Fig. 3.** The $(n_e, T_e)$ plane vs. Fermi degeneracy. Regions are shadowed according to the degree of degeneracy; the higher the darker the greytone. The white region is that investigated in this work.

**Fig. 4.** Proton fusion probability as a function of the incident energy, for different representative values of $n_e$ and $T_e$.

**Fig. 5.** Spectral analysis of scattered protons (a) and multiplication factors (b) for $\alpha$ particles of initial energies $E_{\alpha,I}$ and $E_{\alpha,II}$. The values of $n_e$, $T_e$, $T_i$ and $\gamma$ indicated in panel a) have been used for calculations. Curves in panel a) are given for both the scattering cross sections $\sigma_s$ and $\sigma_R$.

**Fig. 6.** $\alpha$-particle multiplication factor via suprathermal protons as a function of the initial energy, for different values of $T_e$ at a fixed $n_e$ (a), and of $n_e$ at a fixed $T_e$ (b). In a), $T_e$ values (in keV) are indicated next to the curves; as a term of comparison, a curve based only on the Rutherford $\alpha$-$p$ scattering (dashed line) has been generated at $T_e$ = 50 keV.

**Fig. 7.** Multiplication factors $k_{\alpha p}$ and $k_\infty$ as a function of the boron-to-hydrogen ion concentration. Calculations are based on the two representative $\alpha$-particle energies $E_{\alpha,I}$ and $E_{\alpha,II}$, and the indicated values of $n_e$, $T_e$ and $T_i$.

**Fig. 8.** Suprathermal-to-thermonuclear energy ratio by the effect of a weak chain reaction, plotted as a function of $k_\infty$ for several orders of magnitude of the parameter $\tau_c/\tau_\alpha$.

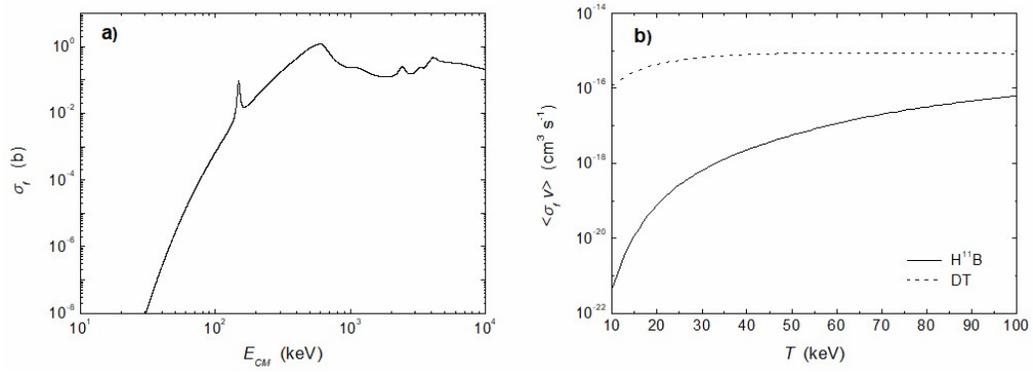

Fig. 1 – F. Belloni

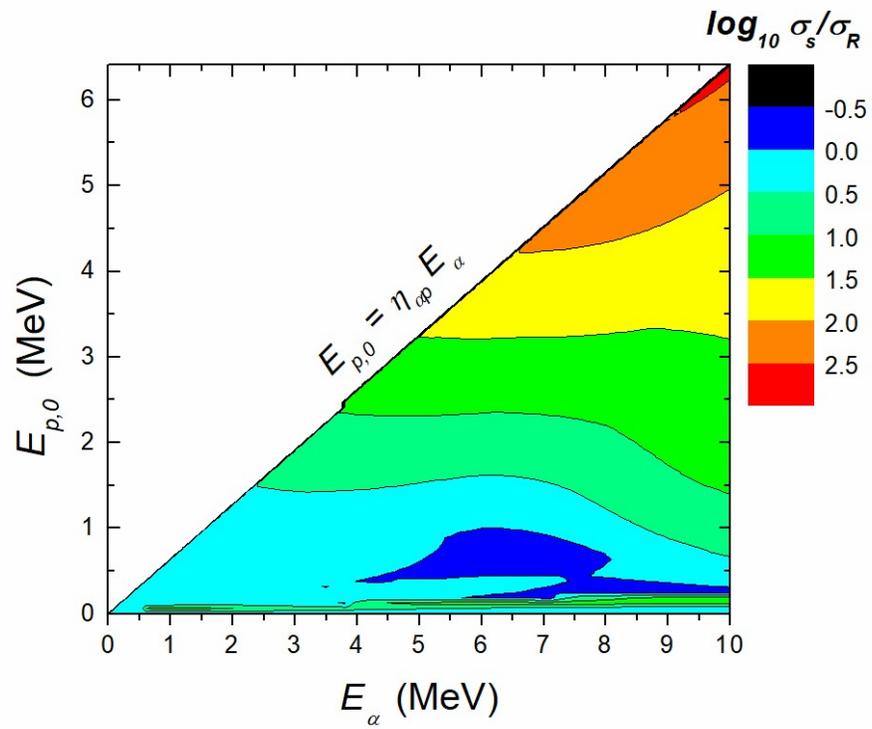

Fig. 2 – F. Belloni

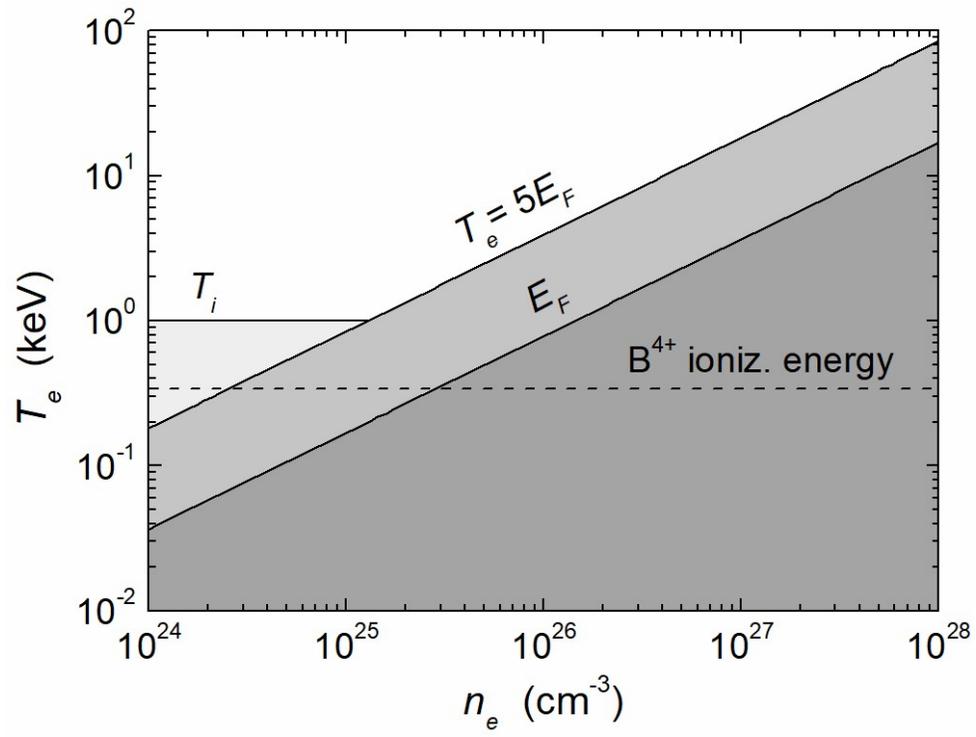

Fig. 3 – F. Belloni

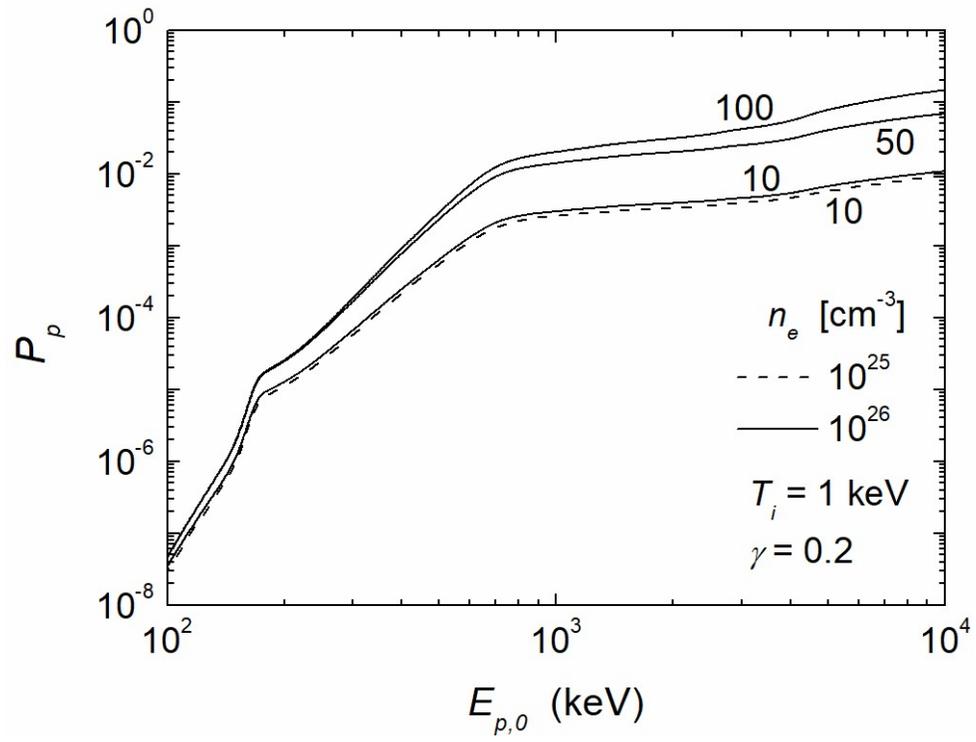

Fig. 4 – F. Belloni

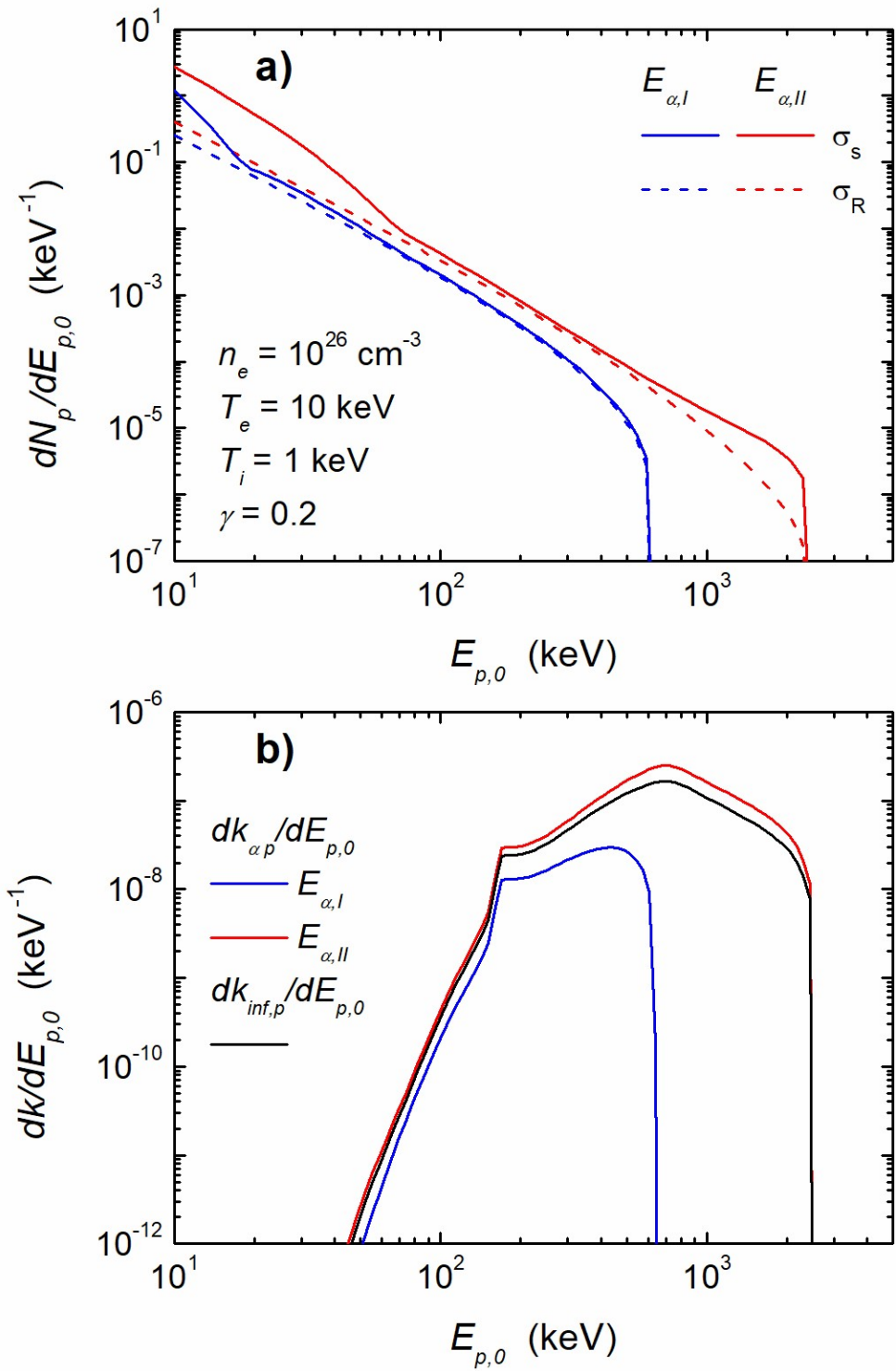

**Fig. 5 – F. Belloni**

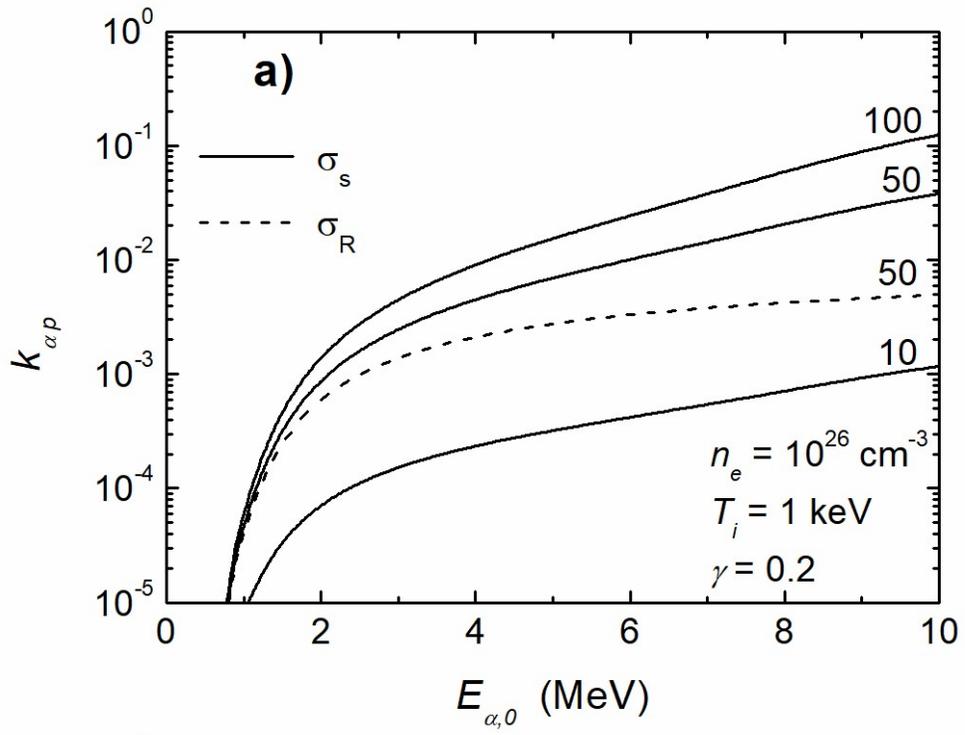

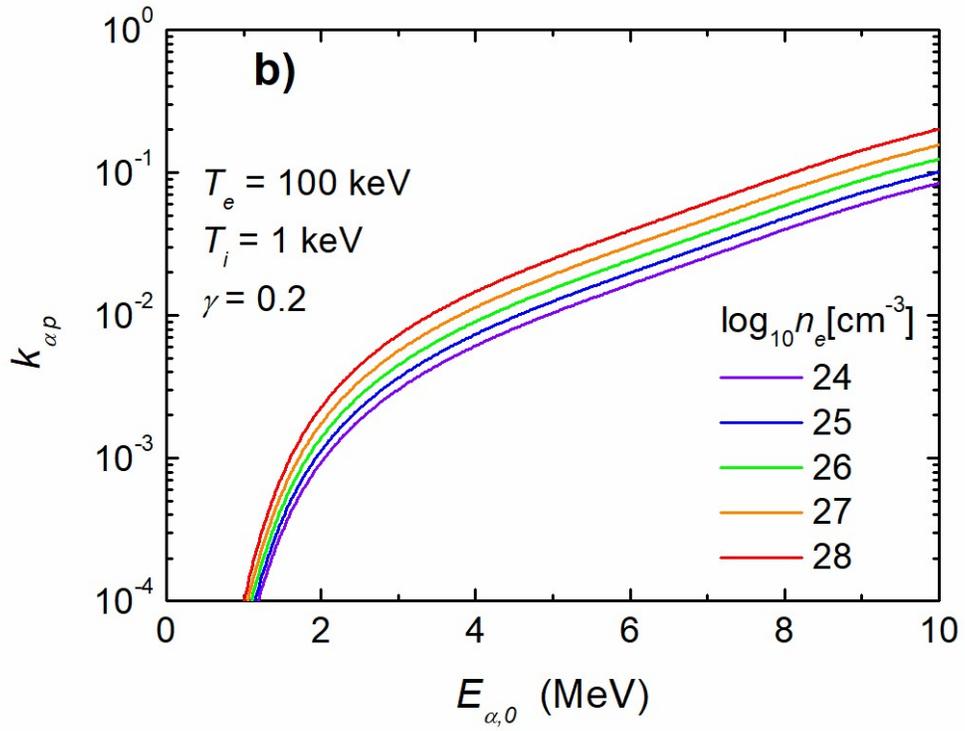

Fig. 6 – F. Belloni

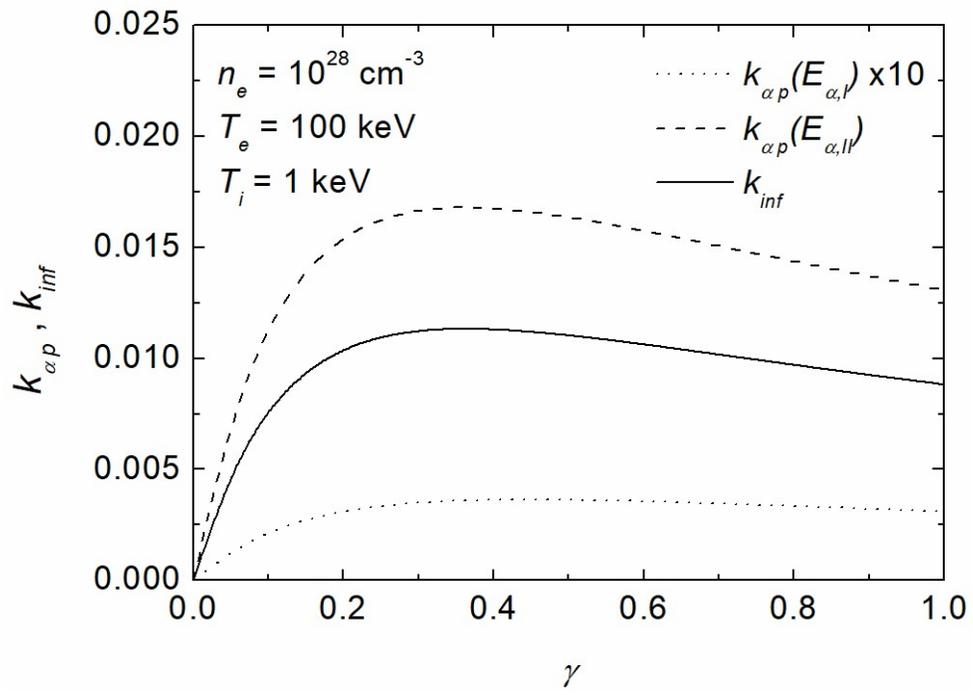

Fig. 7 – F. Belloni

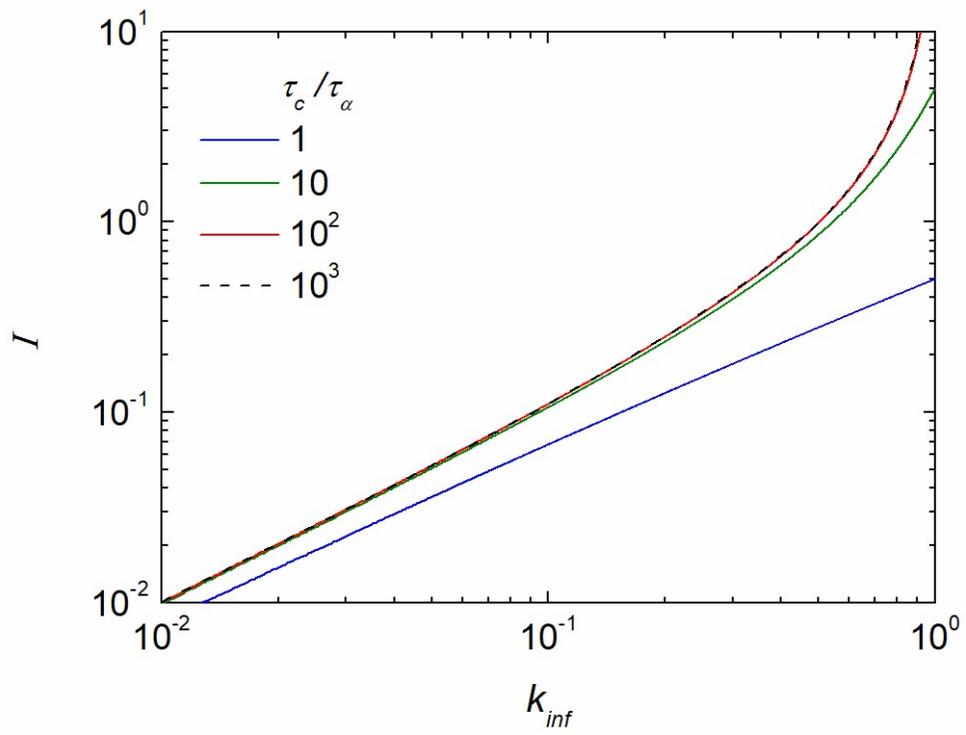

Fig. 8 – F. Belloni